\documentclass{moriond}

\usepackage{amsmath,amssymb}

\def\be{\begin{equation}}
\def\ee{\end{equation}}
\def\bea{\begin{eqnarray}}
\def\eea{\end{eqnarray}}

\newcommand{\Photo}{}

\begin{document}
\vspace*{4cm}
\title{Landscape of Spontaneous CP Violation}

\author{ Yuichiro Nakai}

\address{Tsung-Dao Lee Institute, Shanghai Jiao Tong University, \\
No.~1 Lisuo Road, Pudong New Area, Shanghai, 201210, China \\[1ex]
School of Physics and Astronomy, Shanghai Jiao Tong University, \\
800 Dongchuan Road, Shanghai, 200240, China}

\maketitle\abstracts{
Spontaneous CP violation (SCPV) provides a promising solution to the strong CP problem, explaining the smallness of the QCD $\theta$-angle while generating the Cabibbo-Kobayashi-Maskawa (CKM) phase. In the present work, we review and discuss the realization of SCPV in the supersymmetric framework, which addresses critical issues such as the naturalness of the scale of SCPV and the presence of problematic higher dimensional operators and radiative corrections spoiling the mechanism. It is explicitly shown that SCPV is realized along flat directions and stabilized through supersymmetry-breaking effects and a non-perturbative dynamics, predicting light SCPV sector particles feebly coupled to the Standard Model particles. Furthermore, we discuss the issue of baryon asymmetric Universe in the SCPV framework and point out that the Affleck-Dine mechanism can successfully generate the observed baryon asymmetry with a low reheating temperature compatible with the gravitino dark matter. Our framework predicts a nonzero neutron electric dipole moment which is within the reach of near-future experiments.}

\section{Introduction}

CP symmetry and its violation stand at the heart of modern particle physics. The Standard Model (SM) raises the question of why the strong interaction respects CP symmetry while the weak interaction does not, which is called the strong CP problem.
Looking at our Universe, matter-antimatter asymmetry is a big question to be addressed but its generation requires CP violation.
Furthermore, physics beyond the Standard Model generally contains a new CP violation source which can be probed through
{\it e.g.} measurements of electric dipole moments. In this work, we mostly focus on the strong CP problem and briefly discuss the generation of baryon asymmetry.

The strong CP problem originates from the fact that Quantum Chromodynamics (QCD) allows a CP-violating term in the Lagrangian, 
\begin{equation}
\mathcal{L}_{\theta}
=
\theta \, \frac{g_s^2}{32\pi^2}
G^{a\mu\nu}\tilde{G}^a_{\mu\nu} \ ,
\end{equation}
where $a$ is the color index, $g_s$ is the QCD gauge coupling, $G^{a\mu\nu}$ denotes the gluon field strength and $\tilde{G}^a_{\mu\nu}$ is its dual.
In addition to this term, CP-violating phases arise from the quark mass matrices. The physical CP-violating parameter is then defined as 
\begin{equation}
\bar{\theta}
=
\theta
-
\arg \det (M_u M_d) \ ,
\end{equation}
where $M_u$ and $M_d$ represent  the up and down-type quark mass matrices.
A nonzero $\bar{\theta}$ leads to {\it e.g.}  the neutron electric dipole moment (EDM) whose current experimental upper limit 
imposes a stringent constraint,
\begin{equation}
|\bar{\theta}| \lesssim 10^{-10} \ ,
\end{equation}
which is highly unnatural from a theoretical point of view, as $\bar{\theta}$ is expected to be of order unity in the absence of a symmetry or dynamical mechanism. This discrepancy constitutes the strong CP problem.

The most popular approach to the strong CP problem is the Peccei-Quinn mechanism
\cite{Peccei:1977hh}, in which $\theta$ is promoted to a dynamical field through the introduction of a pseudo-Nambu-Goldstone boson, called the axion
\cite{Weinberg:1977ma,Wilczek:1977pj}. Here, the effective interaction is given by
\begin{equation}
\mathcal{L}
=
\left(
\theta + \frac{a}{f_a}
\right)
\frac{g_s^2}{32\pi^2}
G^{a\mu\nu}\tilde{G}^a_{\mu\nu} \ ,
\end{equation}
with the axion decay constant $f_a$, and non-perturbative QCD effects generate a potential for the axion that dynamically relaxes the combination $\theta + a/f_a$ to zero.
Despite its elegance, the axion solution requires an extremely high-quality global symmetry, which is unlikely to be preserved against quantum gravity effects. This leads to the so-called axion quality problem and motivates the exploration of alternative solutions such as spontaneous CP violation (SCPV), where CP is an exact symmetry of the Lagrangian and is broken only spontaneously.

\section{Spontaneous CP Violation}

In the SCPV framework, the QCD $\theta$ parameter is absent at the Lagrangian level, and a nonzero CP phase arises only through complex vacuum expectation values (VEVs) of scalar fields.
A well-known realization is the Nelson-Barr (NB) mechanism
\cite{Nelson:1983zb,Barr:1984qx,Barr:1984fh}.
A minimal setup includes a vector-like pair of quarks $q, \bar q$, and scalar fields $\eta_\alpha$ whose VEVs break CP spontaneously
\cite{Bento:1991ez}. The relevant interactions are given by
\begin{equation}
\label{NB}
\mathcal{L}
\supset  \mu \bar q q
+ a_{\alpha j} \eta_{\alpha} \bar d_j q
+ y_{ij} H Q_i \bar{d}_j \ ,
\end{equation}
where $H$ is the Higgs doublet.
With these interactions, complex phases of $\langle \eta_\alpha \rangle$ are transmitted to the SM through mixing between the vector-like quarks and the ordinary quarks. The full down-type quark mass matrix takes a block form,
\begin{equation}
\mathcal{M} =
\begin{pmatrix}
\mu & a_{\alpha j} \eta_\alpha \\
0 & m_d
\end{pmatrix}.
\end{equation}
A key property of the NB construction is that the determinant of the above mass matrix is real,
$\arg \det \mathcal{M}  = 0$,
which ensures 
$\bar{\theta}= 0$,
even though CP is broken and a nonzero CKM phase is generated.

Despite the appealing feature, SCPV faces several nontrivial challenges. One of the issues is the stability of a CP-violating vacuum. The scale of SCPV must be hierarchically smaller than the Planck scale, which requires significant fine-tuning as in the case of the electroweak naturalness problem. Another issue is the sensitivity to higher-dimensional operators and radiative corrections, which is highlighted by the following operators:
\begin{equation}
\mathcal{L}\supset
\frac{\eta_\beta^*}{\Lambda} \eta_\alpha \bar q q
+
\frac{\eta_\alpha^*}{\Lambda} H Q\bar q
+
\gamma_{\alpha \beta}\eta_\alpha^*\eta_\beta H^\dagger H \ .
\end{equation}
The first two terms of higher-dimensional operators generically induce complex corrections to the quark mass matrix and thus generate a nonzero $\bar{\theta}$.
The third term can regenerate a strong CP phase through a one-loop process
\cite{Dine:2015jga}.

Supersymmetry (SUSY) offers a natural framework to address the above difficulties. 
SUSY stabilizes the SCPV scale in much the same way it stabilizes the electroweak scale. In addition, SUSY can forbid or strongly suppress dangerous higher-dimensional operators and radiative corrections.
Therefore, it is reasonable to consider SCPV in the supersymmetric framework.\footnote{For non-SUSY approach, see {\it e.g.} the paper by Girmohanta {\it et al}
\cite{Girmohanta:2022giy}.
Besides the NB model, the Hiller-Schmaltz mechanism for SCPV
\cite{Hiller:2001qg,Hiller:2002um} nicely utilizes SUSY properties while the issue of a large Yukawa coupling required for the correct CKM phase in this mechanism can be addressed in a SUSY QCD model
\cite{Nakagawa:2024ddd}.}
In a supersymmetric theory, the physical strong CP phase is given by
\begin{equation}
\bar{\theta} = \theta
-
\arg \det (M_u M_d) - 3 \arg (m_{\tilde g}) \ ,
\end{equation}
with the gluino mass $m_{\tilde g}$.
Now, setting the cosmological constant to zero relates the expectation value of the superpotential with the gravitino mass, $\langle W \rangle \sim m_{3/2} M_{\rm Pl}^2$.
Here, $\langle W \rangle$ generally has a complex phase and leads to a gluino mass phase via the one-loop anomaly mediation effect. As a result, the gravitino mass must be sufficiently small compared with the gluino mass in order not to spoil the solution to the strong CP problem. The corresponding constraint is
\begin{equation}
\label{anomalymed}
\frac{\alpha_s}{4\pi}\frac{m_{3/2}}{m_{\tilde g}} < 10^{-10} \ .
\end{equation}
This is one of the characteristic constraints in the supersymmetric realization of SCPV,
which will be important in a later discussion.
A light gravitino mass indicates that soft SUSY breaking in the Minimal Supersymmetric Standard Model (MSSM) 
is generated by gauge-mediated SUSY breaking.

\section{SCPV via SUSY breaking}

In supersymmetric theories, flat directions, valleys in field space along which the scalar potential is exactly flat, are ubiquitous.
Such flat directions naturally contain a point to spontaneously break CP.
For the stabilization of a CP-violating vacuum by generating positive masses for all scalar fields around the minimum,
there are two qualitatively different possibilities. One is to stabilize the minimum in a purely supersymmetric way.
To the best of our knowledge, this case does not lead to any low-energy degrees of freedom.
The other possibility is to rely on SUSY breaking effects.
In this work, we focus on the latter possibility, where SUSY breaking terms play a crucial role in determining the vacuum structure
\cite{Liu:2025ycm}. This approach has an important feature: it leads to light modes that can be probed experimentally.

Let us consider a superpotential,
\begin{equation}
W = \lambda X(\phi_1 \phi_2 - v^2) \ ,
\end{equation}
where $\phi_1, \phi_2, X$ are chiral superfields and $\lambda$ denotes a real coupling constant.
The corresponding $F$-term scalar potential is given by
\begin{equation}
V_F
=
\lambda^2 |\phi_1 \phi_2 - v^2|^2
+
\lambda^2 |X|^2 (|\phi_1|^2 + |\phi_2|^2) \ .
\end{equation}
This potential is minimized at
\begin{equation}
\langle X \rangle = 0, \qquad
\langle \phi_1 \rangle = v_1 e^{i\theta}, \qquad
\langle \phi_2 \rangle = v_2 e^{-i\theta}, \qquad
v_1 v_2 = v^2,
\end{equation}
which exhibit flat directions parameterized by $v_1 $ (or $v_2$) and  $\theta$.
Expanding the scalar fields $\phi_1, \phi_2$ into radial and phase modes as
\begin{align}
\phi_1(x) &=
\left(v_1 + \frac{\sigma_1(x)}{\sqrt{2}}\right)
\exp\left[i\left(\theta + \frac{\pi_1(x)}{\sqrt{2} v_1}\right)\right], \\[1ex]
\phi_2(x) &=
\left(v_2 + \frac{\sigma_2(x)}{\sqrt{2}}\right)
\exp\left[i\left(-\theta + \frac{\pi_2(x)}{\sqrt{2} v_2}\right)\right],
\end{align}
the massless scalar modes corresponding to the flat directions can be identified as
\begin{align}
\label{massless}
s(x) = \frac{1}{f_a}\left(v_1 \sigma_1(x) - v_2 \sigma_2 (x) \right), \qquad 
a(x) = \frac{1}{f_a}\left(v_1 \pi_1(x) - v_2 \pi_2 (x) \right),
\end{align}
where $f_a \equiv \sqrt{v_1^2 + v_2^2}$.

In order to stabilize the flat directions at a minimum with complex VEVs, more contributions to the scalar potential are needed.
A nontrivial minimum at $\theta \neq 0$ requires at least two terms with different periodicity, {\it e.g.}
$V = C \cos\theta + D \cos(2\theta)$.
Here, we consider two kinds of contributions, one of which originates from SUSY breaking.
The relevant soft SUSY breaking terms are 
\begin{equation}
V_{\rm soft}
=
\left(
\frac{1}{2} b_1 \phi_1^2 + \frac{1}{2} b_2 \phi_2^2 + \mathrm{h.c.}
\right)
+
m_1^2 \phi_1^* \phi_1
+
m_2^2 \phi_2^* \phi_2 \ ,
\end{equation}
among which the first $b$-terms depend on $\theta$. The stability of the scalar potential requires
$|b_1| \le m_1^2 , \  |b_2| \le m_2^2$.
The other contribution is generated by non-perturbative effects of a $SU(N)$ dark supersymmetric QCD with (s)quarks coupled to $\phi_1, \phi_2$. 
The dynamically generated potential is given by
\begin{equation}
V_{\rm dyn}
=
\frac{(\kappa_1^2 + \kappa_2^2)\Lambda^{6-\frac{2}{N}}}
{|\kappa_1 \phi_1 + \kappa_2 \phi_2|^{2-\frac{2}{N}}} \ ,
\end{equation}
where $\kappa_{1,2}$ are real (superpotential) coupling constants of quarks with $\phi_1, \phi_2$ and $\Lambda$ denotes a dark QCD dynamical scale. 
The total scalar potential is then given by
\begin{align}
V_{\rm tot}
&= V_F + V_{\rm soft} + V_{\rm dyn} \nonumber \\
&=
b_1\left(v_1^2 + \frac{b_2}{b_1} \frac{v^4}{v_1^2}\right)\cos(2\theta)
+
m_1^2\left(v_1^2 + \frac{m_2^2}{m_1^2} \frac{v^4}{v_1^2}\right) 
 +
\frac{(\kappa_1^2 + \kappa_2^2)\Lambda^{6-\frac{2}{N}}}
{\left[
\kappa_1^2 v_1^2 + \kappa_2^2 v^4/v_1^2 + 2\kappa_1\kappa_2 v^2 \cos(2\theta)
\right]^{1-\frac{1}{N}}} \ .
\end{align}
With appropriate parameter choices, this potential stabilizes $\theta$ at a nontrivial value ($\neq 0, \pi$) as well as $v_1$.
Note that the vacuum stabilization is realized for the soft SUSY breaking scale $m_{\rm soft}$ hierarchically smaller than the SCPV scale $v$.
The scalar modes in Eq.~\eqref{massless} now get nonzero masses which are controlled by the soft SUSY breaking scale,
\begin{equation}
m_a^2 \sim m_s^2 \lesssim m_{\rm soft}^2 \ ,
\end{equation}
indicating that these states are hierarchically lighter than the SCPV scale.

\begin{figure}
\centering
\includegraphics[width=0.45\linewidth]{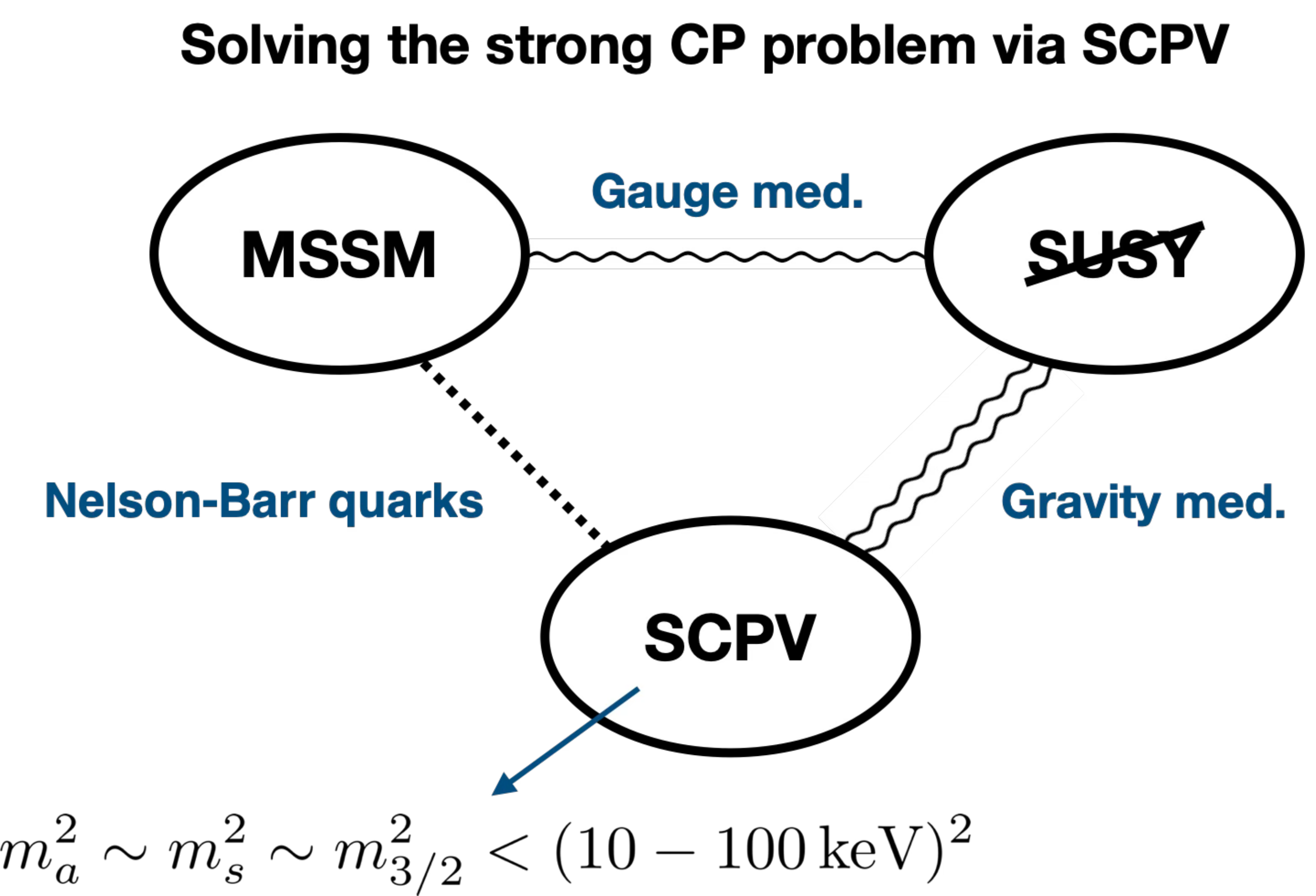}
\caption[]{The setup to solve the strong CP problem via SCPV predicts light particles feebly interacting with the SM.}
\label{fig:light}
\end{figure}

Figure~\ref{fig:light} describes the setup to solve the strong CP problem through SCPV.
The soft SUSY breaking in the MSSM generated by gauge mediation must be larger than the TeV scale.
On the other hand, without any additional structure, the soft SUSY breaking in the SCPV sector is dominantly generated by the gravity mediation effect, so that $m_{\rm soft} \sim m_{3/2}$.
Then, the constraint \eqref{anomalymed} leads to the prediction of light particles with $m_a^2 \sim m_s^2 \lesssim (10 - 100 \, \rm keV)^2$. Interestingly, these light particles are feebly interacting with the SM particles because the NB heavy quarks connect the SCPV sector with the visible MSSM sector as in Eq.~\eqref{NB}.
Detailed phenomenology and cosmology of these light particles will be explored in a future study.

%\begin{figure}
%\centering
%\includegraphics[width=0.3\linewidth]{beta0.5gamma3N3.png}
%\caption[]{same figure with draft option (left), normal (center) and rotated (right)}
%\label{fig:parameter space}
%\end{figure}

%For convenience, we introduce the dimensionless parameters
%\begin{equation}
%\alpha_b = \frac{b_2}{b_1}, \qquad
%\alpha_m = \frac{m_2^2}{m_1^2}, \qquad
%\beta = \frac{b_1}{m_1^2}, \qquad
%\gamma = \frac{\Lambda^{6-\frac{2}{N}}}{m_1^2 v^{4-\frac{2}{N}}},
%\end{equation}
%which characterize the structure of the scalar potential and determine the vacuum configuration.

\section{Baryogenesis}

For considering the baryon asymmetric Universe in the context of SCPV, one naturally wonders the origin of CP violation for baryogenesis. Moreover, the conventional thermal leptogenesis requires a high reheating temperature, leading to the overproduction of gravitinos, and is inconsistent with our supersymmetric framework.

In this context, the Affleck-Dine (AD) mechanism
\cite{Affleck:1984fy} offers a natural and viable scenario of baryogenesis and is compatible with a low reheating temperature. The AD mechanism utilizes a flat direction, parameterized by a complex scalar field $\phi = \frac{\varphi}{\sqrt{2}}\, e^{i\theta}$ carrying baryon or lepton number. Here, this AD field can be identified with
\cite{Fujikura:2022sot}
\begin{equation}
\phi^3 \sim Q \bar{q} L \ ,
\end{equation}
including the NB (s)quark $\bar{q}$.
During inflation, the field $\phi$ develops a large VEV. After the onset of oscillation, it acquires a coherent rotational motion in the complex plane, which generates $B-L$ number density given by
$n_{B-L} = q_{B-L}\,\varphi^2\,\dot{\theta}$,
where $q_{B-L}$ is the $B-L$ charge of the flat direction.
Note that explicit CP violation in the Lagrangian is not required. Instead, the necessary CP violation is provided by the initial condition of the scalar field evolution, making the mechanism particularly compatible with the SCPV framework.

\begin{figure}
\centering
\includegraphics[width=0.5\linewidth]{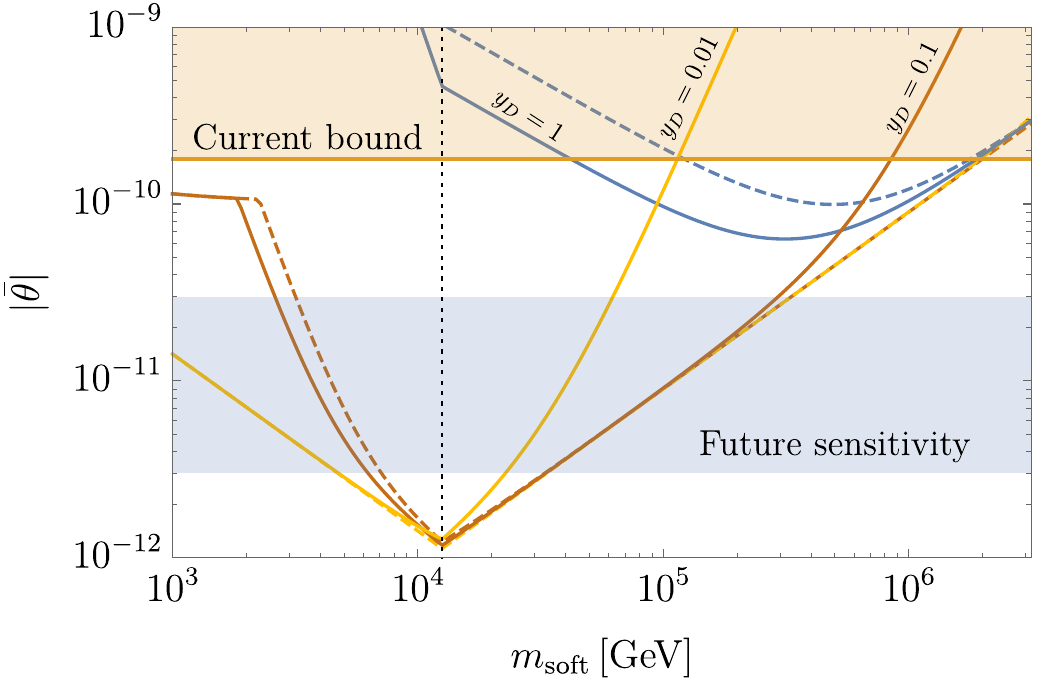}
\caption[]{The induced strong CP phase $\bar{\theta}$ as a function of the soft SUSY breaking scale of the MSSM sector
(solid and dashed curves)
\cite{Fujikura:2022sot}. The orange region is excluded by the current upper limit on the neutron EDM.
The blue band shows its future sensitivity.
The SCPV scale and reheating temperature are chosen to obtain the observed asymmetry and DM 
(the consistency with the SCPV vacuum stabilization via SUSY breaking will be explored in a future study).}
\label{fig:parameter space}
\end{figure}

When the gravitino constitutes dark matter (DM), the Lyman-$\alpha$ bound implies
\begin{equation}
m_{3/2} \gtrsim 5.3 \,\mathrm{keV} .
\end{equation}
There exists a viable parameter region where both the observed baryon asymmetry and DM abundance are realized
\cite{Fujikura:2022sot}.
A striking implication of this framework is that the strong CP phase induced by {\it e.g.} the anomaly mediation effect is small but nonzero, leading to a neutron EDM within the reach of near-future experiments, as shown in figure~\ref{fig:parameter space}.

\section{Conclusion}

In the present work, we have explored a framework based on SCPV as a viable solution to the strong CP problem. It was discussed that such a framework can be consistently realized in a supersymmetric theory.
We have shown that a SCPV vacuum stabilization via SUSY breaking predicts
light particles feebly interacting with the SM particles.
The cosmological implications were also explored. The AD mechanism provides a natural realization of baryogenesis compatible with a low reheating temperature, which makes it possible to avoid the gravitino overproduction problem. The parameter space consistent with the observed baryon asymmetry and gravitino DM leads to a neutron EDM within the reach of near-future experiments.
Overall, SCPV offers a viable and testable framework that connects the strong CP problem, baryogenesis and DM.

\section*{Acknowledgments}

The author would like to thank the organizers of the 60th Rencontres de Moriond, the session of Electroweak Interactions and Unified Theories for putting together such a great conference. He is also grateful to all the collaborators of the project, Kohei Fujikura, Sudhakantha Girmohanta, Seung J. Lee, Fangchao Liu, Shota Nakagawa, Ryosuke Sato, Motoo Suzuki, Yaoduo Wang and Masaki Yamada.
He is supported by Natural Science Foundation of Shanghai.

\section*{References}
\bibliography{moriond}

\end{document}